\documentclass[aps,prd,preprint,superscriptaddress,preprintnumbers,amsmath]{revtex4-2}

\usepackage{mlmodern}
\usepackage{newpxtext}

\usepackage{graphicx}
\usepackage{braket}
\usepackage{lipsum}
\usepackage{subfig}
\usepackage{hyperref}
\usepackage{ulem}

\usepackage{color}

\begin{document}
\preprint{RIKEN-iTHEMS-Report-26}

\title{$\bar{D}$-meson Nucleon Scattering from Lattice QCD at the Physical Point}

\author{Wren~Yamada}
\email{rento.yamada@riken.jp}
\affiliation{RIKEN Center for Interdisciplinary Theoretical and
Mathematical Sciences (iTHEMS), RIKEN, Wako 351-0198, Japan}

\author{Yan~Lyu}
\affiliation{RIKEN Center for Interdisciplinary Theoretical and
Mathematical Sciences (iTHEMS), RIKEN, Wako 351-0198, Japan}

\author{Kotaro~Murakami}
\affiliation{Department of Physics, Institute of Science Tokyo, 2-12-1 Ookayama, Megro, Tokyo 152-8551, Japan}
\affiliation{RIKEN Center for Interdisciplinary Theoretical and
Mathematical Sciences (iTHEMS), RIKEN, Wako 351-0198, Japan}

\author{Takumi~Doi}
\affiliation{RIKEN Center for Interdisciplinary Theoretical and
Mathematical Sciences (iTHEMS), RIKEN, Wako 351-0198, Japan}

\collaboration{HAL QCD Collaboration}

\date{\today}

\begin{abstract}
We report the first lattice QCD study of the $s$-wave scattering of the $\bar{D}$-meson and the nucleon at the physical point, utilizing (2+1)-flavor configurations generated by the HAL QCD collaboration with a pion mass of $m_\pi\simeq 137$ MeV and a lattice spacing of $a\simeq0.084$ fm. 
By applying the HAL QCD method to the four-point correlation function of the $\bar{D}N$ system, we obtain a leading-order potential of the derivative expansion of the interaction kernel, which is then used to extract the $s$-wave phase shifts of low-energy $\bar{D}N$ scattering.
Both the isospin $I=0$ and $I=1$ channels have a short-range repulsive core and a shallow attractive pocket in the intermediate to long-range region, though the $I=0$ channel is more attractive than the $I=1$ channel.
We also observe that the $\bar{D}N$ potential exhibits more attraction than the $KN$ potential, which is its analog in the strange sector.
In terms of the $s$-wave phase shifts, the $I=0$ channel shows a weak attractive behavior in the low-energy region with a positive scattering length of $ 0.425 \pm 0.163 (_{-0.105}^{+0.129})$ fm, whereas the $I=1$ channel shows repulsion with a negative scattering length of $-0.235 \pm 0.046 (_{-0.007}^{+0.040})$ fm.
No bound states are found in both isospin channels, indicating the absence of a pentaquark state in the $s$-wave $\bar{D}N$ system.
\end{abstract}


\maketitle


\section{Introduction} \label{sec:intro}
The study of hadron-hadron interactions, including those involving charm, is vital in various aspects of hadron and nuclear physics.
For example, it is important in understanding hadron resonances, especially exotic hadrons such as tetraquark or pentaquark states beyond conventional quark configurations (e.g., $X(3872)$, $T_{cc}$, $P_c$), situated near the two-body hadronic thresholds with charmed hadrons \cite{Chen:2016qju,Hosaka:2016pey,Lebed:2016hpi,Esposito:2016noz,Guo:2017jvc,Ali:2017jda,Olsen:2017bmm,Karliner:2017qhf,Liu:2019zoy,Brambilla:2019esw,Chen:2022asf}. 
Also, interactions between charmed hadrons and nucleons give us insight into the spectrum of charmed nuclei, and in-medium modifications of mass and width of charmed hadrons in nuclear matter \cite{Hosaka:2016ypm,Krein:2017usp,Brodsky:1989jd,Tsushima:1998ru,Hayashigaki:1998ey,Hayashigaki:2000es,Tolos:2004yg,Lutz:2005vx}.

In this paper, we investigate the interaction between the $\bar{D}$ meson and the nucleon ($N$) at low energies using first-principles lattice QCD calculations of the underlying theory of quantum chromodynamics (QCD). 
The $\bar{D}N$ system is the charm analog of the $KN$ system, with a quark configuration of $\bar{c}qqqq$ ($q=u$ or $d$) with no quark-antiquark annihilation of the valence quarks. 
Thus, the $\bar{D}$ meson and the nucleon can form an explicitly exotic pentaquark state if the attraction between the two is strong enough to form a bound state.
It is discussed in some literature \cite{Hosaka:2016ypm,Yasui:2009bz,Gamermann:2010zz,Yamaguchi:2011xb,Yamaguchi:2022oqz} that heavy-quark spin symmetry \cite{Isgur:1989vq} plays an important role in yielding attraction between the $\bar{D}$ meson and the nucleon.
In contrast to the case of the kaon, the approximate degeneracy of the $\bar{D}$ meson and the $\bar{D}^\ast$ meson, with a mass difference of 140 MeV, is expected to enhance the coupling between the $\bar{D}N$ and $\bar{D}^\ast N$ channels, effectively generating more attraction. 
There have been various theoretical studies using effective models to investigate the $\bar{D}N$ interaction~\cite{Hofmann:2005sw,Haidenbauer:2007jq,Yasui:2009bz,Gamermann:2010zz,Yamaguchi:2011xb,Carames:2012bd,Fontoura:2012mz,Yamaguchi:2022oqz} that disagree qualitatively.
While some theoretical models indicate a strong attraction between the $\bar{D}$ meson and the nucleon sufficient to form bound states \cite{Gamermann:2010zz,Yasui:2009bz,Yamaguchi:2011xb,Yamaguchi:2022oqz}, other models only show weak attraction or repulsion. 
The scattering lengths also vary depending on the model
likely due to the undetermined model parameters.
From the experimental side, conventional two-body scattering experiments are incredibly challenging in the case of $\bar{D}N$ system.
Femtoscopy analysis of the $D^- p$ momentum correlation function from $pp$ collisions by the ALICE collaboration~\cite{ALICE:2022enj} is the first experimental measurement concerning the two-body interaction between open charmed mesons and nucleons. 
Though present data do not yield a definitive conclusion on the low-energy scattering behavior of $\bar{D}N$, the LHC RUN-3 data with significantly larger statistics is expected to be available in the near future, which should improve our understanding of the $\bar{D}N$ system.
Given these circumstances, we anticipate that a realistic lattice QCD investigation will play a critical role in both theoretical and experimental studies concerning the $\bar{D}N$ system at low energies.

The HAL QCD method~\cite{Ishii:2006ec,Aoki:2009ji,Ishii:2012ssm,Aoki:2012tk} is one of the established ways to extract hadron-hadron interactions from lattice QCD simulations.
While the L\"uscher's finite volume method \cite{Luscher:1986pf,Luscher:1990ux} extracts scattering information from the volume-dependence of the energy shifts in finite volume, the HAL QCD method extracts interaction information from the temporal and spatial profiles of the Nambu-Bethe-Salpeter (NBS) correlation functions.
Many studies of hadron-hadron interactions using the HAL QCD method have been conducted employing the configurations near the physical point with a pion mass of $m_\pi=146$ MeV~\cite{Doi:2017zov,Gongyo:2017fjb,HALQCD:2018qyu,HALQCD:2019wsz,Lyu:2021qsh,Lyu:2022imf,Lyu:2023xro,Lyu:2024ttm}.
Now, it is possible to perform physical-point simulations using the new configurations generated by the HAL QCD collaboration with a physical pion mass of $m_\pi = 137$ MeV~\cite{Aoyama:2024cko}. 
Employing this physical-point configuration, the $N\Omega_{ccc}$ system \cite{Zhang:2025zaa} and $KN$ system \cite{Murakami:2025owk} have been studied recently.
In this study, we investigate the $\bar{D}N$ system using the same configuration within a single-channel framework,
in which the off-shell effects of the $\bar{D}^\ast N$ channel are implicitly incorporated. 
We determine the interaction potential, $s$-wave phase shifts, scattering length and effective range.
We note that this is the first investigation of the $\bar{D}N$ system from lattice QCD simulations.

The paper is organized as follows. 
In Sec.\,\ref{sec:method}, we introduce the time-dependent HAL QCD method applied to a meson-baryon system. 
We outline how to extract the potential and phase shifts from correlation functions.
In Sec.\,\ref{sec:setup}, details of the lattice simulation setup are presented. 
Then in Sections \,\ref{sec:lopot} and \ref{sec:scattpar}, the results of our HAL QCD analysis are given. 
In Sec.\,\ref{sec:lopot}, we present the leading-order potentials in the derivative expansion of $\bar{D}N$ in the isospin $I=0$ and $I=1$ channels. We also compare the $\bar{D}N$ potentials to the $KN$ potentials presented in Ref.\,\cite{Murakami:2025owk}.
In Sec.\,\ref{sec:scattpar}, we present the $s$-wave phase shifts, scattering length and effective range of the $\bar{D}N$ system. 
Section \ref{sec:conclusion} concludes this paper.

\section{HAL QCD Method} \label{sec:method}
Here we summarize the HAL QCD method~\cite{Ishii:2006ec,Aoki:2009ji,Ishii:2012ssm,Aoki:2012tk} to outline the procedure in obtaining the potentials and phase shifts of $\bar{D}N$ scattering from lattice QCD simulations. 
We will consider the general case of a scattering problem of a meson and a baryon. 

Let us define the R-correlator $R(\vec{r},t)$ as
\begin{equation}
    R(\vec{r},t) =
    \frac{\sum_{\vec{x}} \braket{0|M(\vec{x}+\vec{r},t)B(\vec{x},t)\bar{\mathcal{J}}(0)|0}}{C^{M}_{2\text{pt}}(t)\,C^{B}_{2\text{pt}}(t)}, \label{eq:rcorr}
\end{equation}
where $C^{M (B)}_{2\text{pt}}(t)$ denotes the two-point correlation function of the meson (baryon) operator at Euclidean-time $t$.
$M$ ($B$) represents the point-like sink operators for the meson (baryon) field, and operator $\bar{\mathcal{J}}(0)$ is a source operator placed at $t=0$, overlapping with the meson-baryon two-body states. 

For sufficiently large $t$, $R(\vec{r},t)$ behaves as
\begin{equation}
    R(\vec{r},t) = \sum_n a_n \psi_n(\vec{r}; E_n) e^{-\Delta E_n t}+\mathcal{O}(e^{-\Delta E^\ast t}),
\end{equation}
where $\psi_n(\vec{r}; E_n)$ is the equal-time Nambu-Bethe-Salpeter (NBS) wave function and $a_n = \braket{n|\bar{\mathcal{J}}(0)|0}$ is the overlap of the source operator for the $n$-th eigen state of the meson-baryon system.
$\Delta E_n$ is the eigen energy of the $n$-th eigen state and $\Delta E^\ast$ is the energy of the lowest inelastic scattering states, with respect to the two-body threshold at $E=m_{M}+m_{B}$. 
$m_{M}$ ($m_{B}$) corresponds to the mass of the meson (baryon).
When considering sufficiently large time, contributions from the inelastic states are exponentially suppressed, and can be regarded as negligible.

As shown in Refs.\,\cite{Ishii:2012ssm, Miyamoto:2017tjs,Murakami:2020yzt}, 
the R-correlator satisfies the following equation
\begin{equation}\label{eq:eom}
    \left[\frac{1 + 3\delta^2}{8\mu} \partial_t^2 - \partial_t - H_0 + \mathcal{O}(\delta^2 \partial_t^3)\right] R(\vec{r},t) = \int d^3 r' U(\vec{r},\vec{r'}) R(\vec{r'},t),
\end{equation}
where $H_0 = -\nabla^2 / (2\mu)$. $\mu = (m_M m_B)/(m_M + m_B)$ is the reduced mass of the system, and $\delta = (m_M - m_B)/(m_M + m_B)$ is a dimensionless parameter associated with the difference between the meson and baryon masses.
The interaction kernel, $U(\vec{r},\vec{r'})$, is an energy-independent non-local interaction, which ensures the unitarity of the $S$-matrix and manifests the correct scattering phase shifts up to the lowest inelastic threshold.

In practice, $U(\vec{r},\vec{r'})$ is often expanded by the derivative expansion as
\begin{equation}\label{eq:derivexp}
    U(\vec{r},\vec{r'}) = V_{\text{LO}}(r) \delta^3(\vec{r} - \vec{r'})+\mathcal{O}(\nabla^n).
\end{equation}
By substituting the first term of Eq.\,\eqref{eq:derivexp} into Eq.\,\eqref{eq:eom}, we can obtain the leading-order potential $V_{\text{LO}}(r)$ from the R-correlator as
\begin{equation}\label{eq:LOpot_}
    V_{\text{LO}}(r) = R(\vec{r},t)^{-1}\left[\frac{1 + 3\delta^2}{8\mu}\partial_t^2 - \partial_t - H_0\right] R(\vec{r},t),
\end{equation}
where we omitted terms of $\mathcal{O}(\delta^2 \partial_t^3)$. 
The $\delta^2 \partial_t^3$ term has been numerically checked to be insignificant for our system of interest.
In this paper, we limit our analysis to the leading-order potential $V_{\text{LO}}(r)$. 
Effects from the higher-order terms in the derivative expansion are found to be under control at low energies, as discussed in Appendix \ref{sec:fvenergy}.

We then obtain the phase shifts by solving the Schr\"{o}dinger equation in infinite volume 
\begin{equation}\label{eq:Schrodinger}
    \left(-\frac{\nabla^2}{2\mu}+V_{\text{fit}}(r) \right) \psi(\vec{r}; E) = \frac{k^2}{2\mu}  \psi(\vec{r}; E),
\end{equation}
where $V_{\text{fit}}(r)$ is the potential function fitted to the original potential $V_{\text{LO}}(r)$ and $k$ is the relative momentum in the center-of-mass frame that is related to the energy as $E=\sqrt{k^2+m_M^2}+\sqrt{k^2+m_B^2}$.

\section{Setup} \label{sec:setup}
In this section, we present the details of our lattice simulation.
We use the following operators for the local sink operators of the $\bar{D}$ meson and the nucleon,
\begin{align}
    \bar{D}_i(x) &= \bar{c}(x) i \gamma_5 q_i(x), \label{eq:sink1}\\
    N_{\alpha,i}(x) &= \varepsilon_{abc} \left[u^a(x)^T C \gamma_5 d^b(x)\right] q^c_{\alpha,i}(x),\label{eq:sink2}
\end{align}
where $\bar{D} = (\bar{D}^0, D^-)^T$, $N_{\alpha} = (p_\alpha, n_\alpha)^T$, and $q = (u,d)^T$. 
The indices $a$, $b$, and $c$ denote the color indices, and $\alpha$ denotes the spinor index. 
The charge conjugation operator is given by $C = \gamma_4 \gamma_2$.
For the source operator, we use a wall source operator $\mathcal{J}^{\text{wall}}(t_0)$ with Coulomb gauge fixing, which enhances the overlap with the low-energy $\bar{D}N$ states.
$\mathcal{J}^{\text{wall}}(t_0)$ is explicitly given by
\begin{gather}
    \mathcal{J}^{\text{wall}}(t_0)=\bar{D}^{\text{wall}}(t_0)N^{\text{wall}}(t_0)
\end{gather}
where $\bar{D}^{\text{wall}}$ and $N^{\text{wall}}$ are obtained by replacing the quark fields in Eqs.\,\eqref{eq:sink1} and \eqref{eq:sink2} by
\begin{gather}
    q^{\text{wall}}(t_0) = \sum_{\vec{x}} q(\vec{x},t_0).
\end{gather}

The correlation functions are computed with the unified contraction algorithm~\cite{Doi:2012xd},
using the gauge configurations, "HAL-conf-2023", generated by the HAL QCD collaboration at the physical pion mass \cite{Aoyama:2024cko}. 
(2+1)-flavor non-perturbatively $\mathcal{O}(a)$-improved Wilson quark action with stout smearing, along with the Iwasaki gauge action at a bare coupling constant of $\beta = 1.82$ is used. 
The size of the lattice is $L^4 = 96^4 $ with a lattice spacing of $a = 0.084372(54)(^{+109}_{-6})$ fm, corresponding to an inverse lattice spacing of $a^{-1} = 2338.8(1.5)(^{+0.2}_{-3.0})$ MeV.
The pion and nucleon masses are $m_\pi = 137.1(0.3)(^{+0.0}_{-0.2})$ MeV and $m_N = 939.7(1.8)(^{+0.2}_{-1.7})$ MeV, respectively.
We use this fixed value of $m_N$ for the nucleon mass in our analysis.
The effective mass plot for the nucleon is presented in Ref.\,\cite{Aoyama:2024cko}.
For the charm quark action, we use the relativistic heavy-quark (RHQ) action given in Ref.\,\cite{Aoki:2001ra},
which eliminates 
discretization errors up to next-to-leading-order.
Two sets of parameters for the RHQ action given in Ref.\,\cite{Namekawa:2017XO} are used in this study, where one set (Set-I) with a slightly heavier charm quark mass and another set (Set-II) with a slightly lighter charm quark mass compared to the physical charm quark mass. 
The corresponding masses of the $\bar{D}$ ($\bar{D}^\ast$) meson in Set-I and Set-II are given in Tab.\,\ref{tab:meson_mass} of Appendix \ref{sec:dmeson_mass}.

For statistics, we use 360 gauge configurations sampled at intervals of five trajectories. 
To enhance statistical precision, the correlation functions are averaged over both forward and backward propagation directions for each configuration. 
In addition, $96 \times 4$ measurements are performed by shifting the source operator in the temporal direction and applying four rotations for the temporal direction using the hypercubic symmetry.
The total number of measurements is thus $276,480$,
and these measurements are averaged within each gauge configuration prior to the statistical analysis.
Statistical uncertainties are then estimated by using the jackknife sampling method after binning the configuration-averaged data.
The bin size dependence was checked using
bin sizes of 20, 30, and 40 configurations, and found to be small.
In this paper, we present the results with the bin size fixed to 30 configurations.

Isospin projections onto the isospin $I=0$ and $I=1$ channels are performed by taking the linear combination of the 4-point correlation functions as
\begin{align}
    C^{\bar{D}N(I=0)}_{4\text{pt},\alpha}(\vec{r},t) &= \frac{1}{\sqrt{2}}C^{\bar{D}^{0}n}_{4\text{pt},\alpha}(\vec{r},t)-\frac{1}{\sqrt{2}}C^{D^{-}p}_{4\text{pt},\alpha}(\vec{r},t), \\
    C^{\bar{D}N(I=1)}_{4\text{pt},\alpha}(\vec{r},t) &= \frac{1}{\sqrt{2}}C^{\bar{D}^{0}n}_{4\text{pt},\alpha}(\vec{r},t)+\frac{1}{\sqrt{2}}C^{D^{-}p}_{4\text{pt},\alpha}(\vec{r},t).
\end{align}
Also, since we will only be concerned with the $s$-wave component in this study, the R-correlators are projected onto the $A_1^+$ representation of the octahedral group $\mathcal{O}_h$ as
\begin{equation}
    R_{A_1^+}(\vec{r},t) = \frac{1}{48} \sum_{g \in \mathcal{O}_h} \chi_{A_1^+}(g) R(g^{-1}\vec{r},t),
\end{equation}
where $\chi_{A_1^+}$ are the characters of the $A_1^+$ representation. 
To further suppress contributions from higher partial waves with orbital angular momenta of $l = 4, 6, \dots$, we apply the Misner method discussed in detail in Refs.\,\cite{Misner:1999ab,Miyamoto:2019jjc}.
In the subsequent sections we show the results with $n_{\text{max}} = 2$, $l_{\text{max}} = 4$, and $\Delta/a = 1.0$, following the notation of Ref.\,\cite{Miyamoto:2019jjc}.
The results are stable with respect to the change of $n_{\text{max}}$ and $l_{\text{max}}$ in the region $\Delta/a \approx 1$.

\section{Interaction Potential between $\bar{D}$ and $N$} \label{sec:lopot}
The LO potential $V_{\text{LO}}(r)$ of the $\bar{D}N$ system can be obtained by applying Eq.\,\eqref{eq:LOpot_} to the R-correlators computed under our setup discussed in the previous section.
In this section we show $V_{\text{LO}}(r)$ for three consecutive time slices $t/a = 13$-$15$. 
The window of time slices has been chosen where contributions from inelastic excited states at small times are suppressed while avoiding the exponentially growing statistical errors at larger times. 
In terms of the charm quark mass dependence,
we take a weighted average of the LO potentials calculated for Set-I and Set-II, i.e., $V_{\text{LO}}(r)=w_1 V^{(\textrm{Set-I})}_{\text{LO}}(r)+w_2 V^{(\textrm{Set-II})}_{\text{LO}}(r)$
throughout the rest of the analysis, 
where the weights $(w_1, w_2) = (0.411, 0.589)$ are determined to linearly interpolate the $\bar{D}$ meson masses of Set-I and Set-II to the physical (PDG) value of the $\bar{D}^0$ meson mass. 
A comparison of the LO potentials obtained from Set-I and Set-II is presented in Appendix\,\ref{sec:interpolation}.

\begin{figure*}[!htb] 
    \centering
    \subfloat{\includegraphics[width=0.5\linewidth]{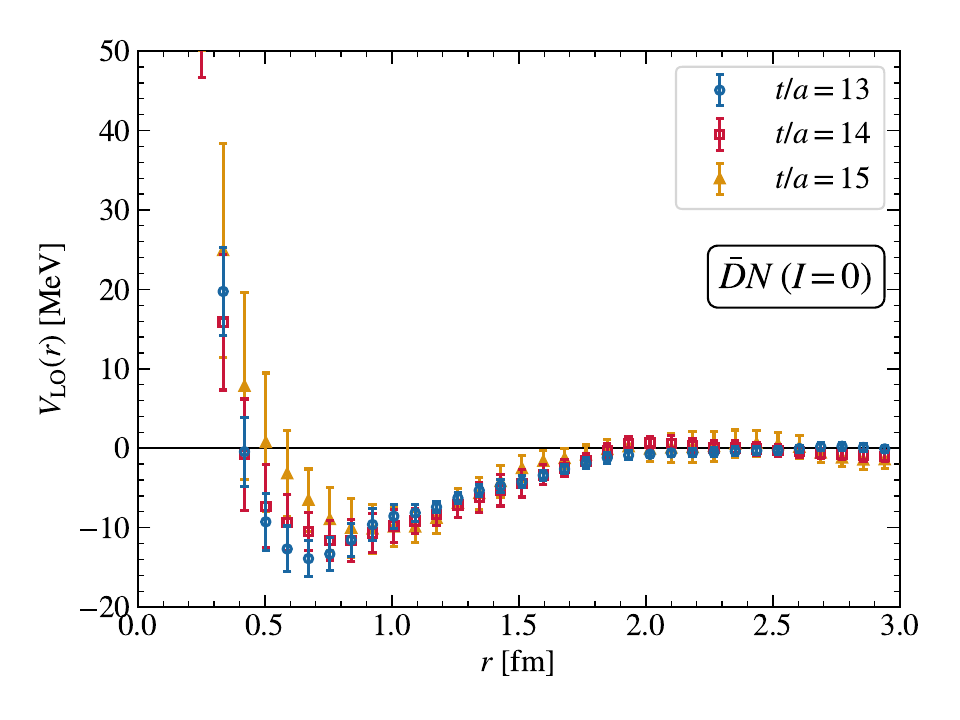}}
    \subfloat{\includegraphics[width=0.5\linewidth]{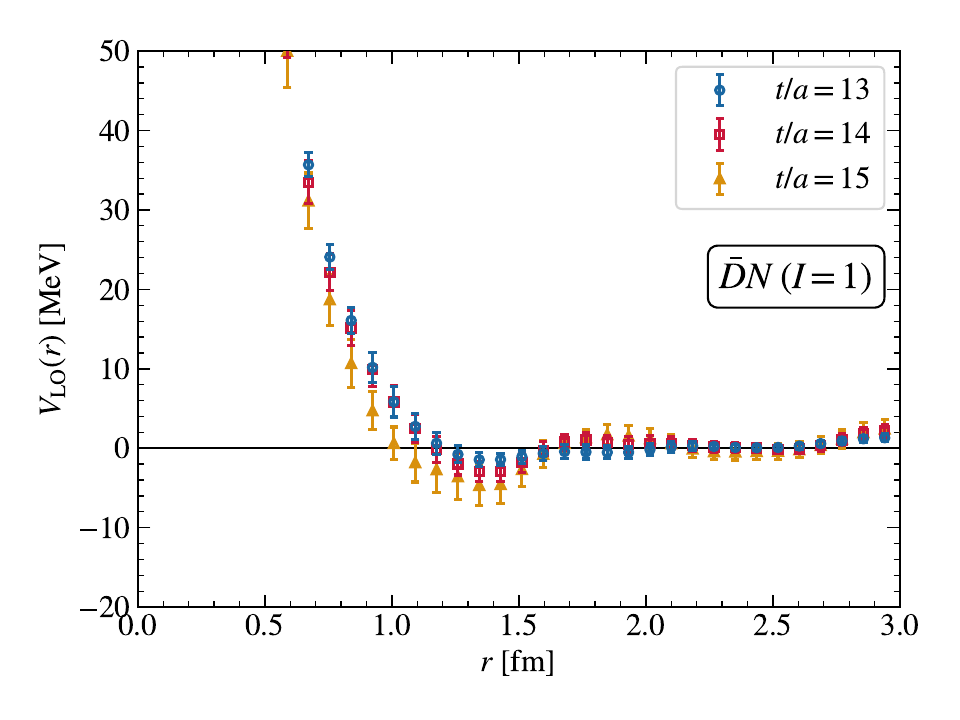}}
    \caption{LO potential $V_{\text{LO}}(r)$ for the $\bar{D}N$ system in the isospin $I=0$ channel (left), and the $I=1$ channel (right) at Euclidean time slices $t/a=13$ (blue, circle), $t/a=14$ (red, square) and $t/a=15$ (gold, triangle). 
    }\label{fig:Lopot_}
\end{figure*}
The LO potentials $V_{\text{LO}}(r)$ for the $I=0$ channel and the $I=1$ channel are shown in Fig.\,\ref{fig:Lopot_}.
$V_{\text{LO}}(r)$ shows little time dependence within the time window of $t/a = 13$-$15$, indicating that contributions from excited states and the truncation error in the derivative expansion are negligible in comparison to the statistical fluctuations. 
We observe that $V_{\text{LO}}(r)$ has a repulsive core in the short-distance region and an attractive pocket in the intermediate-to-long-distance region for both isospin channels.
For the $I=0$ channel, $V_{\text{LO}}(r)$ has a repulsive core in the short-distance region up to around $0.5$ fm and an attractive pocket between $0.5$ fm to $2.0$ fm.
The attractive pocket has a maximum depth of about $10$ MeV at around $r\simeq 1.0$ fm.
For the $I=1$ channel, $V_{\text{LO}}(r)$ has a repulsive core in the short-distance region up to around $1.0$ fm and an attractive pocket between $1.0$ fm to $2.0$ fm.
The attractive pocket has a maximum depth of a $2$-$5$ MeV at around $r\simeq 1.2$ fm.
As a general feature, the $I=1$ channel shows stronger repulsion (or smaller attraction) compared to the $I=0$ channel.

The LO potential can also be decomposed into isospin-independent parts and dependent parts as
\begin{gather}
    V_{\text{LO}}(r)=V_0(r)+V_\tau(r)\vec{\tau}_{\bar{D}}\cdot\vec{\tau}_{N},
\end{gather}
where $\vec{\tau}_{\bar{D}}$ ($\vec{\tau}_{N}$) are the isospin operators acting on the $\bar{D}$ meson (nucleon).
$V_0(r)$ and $V_\tau(r)$ can be expressed by $V_{\text{LO}}^{(I=0,1)}$ as
\begin{align}
    V_0(r)&=\frac{1}{4}V_{\text{LO}}^{(I=0)}+\frac{3}{4}V_{\text{LO}}^{(I=1)},\\
    V_\tau(r)&=-\frac{1}{4}V_{\text{LO}}^{(I=0)}+\frac{1}{4}V_{\text{LO}}^{(I=1)}.
\end{align}
\begin{figure*}[!htb]
    \centering
    \subfloat{\includegraphics[width=0.50\linewidth]{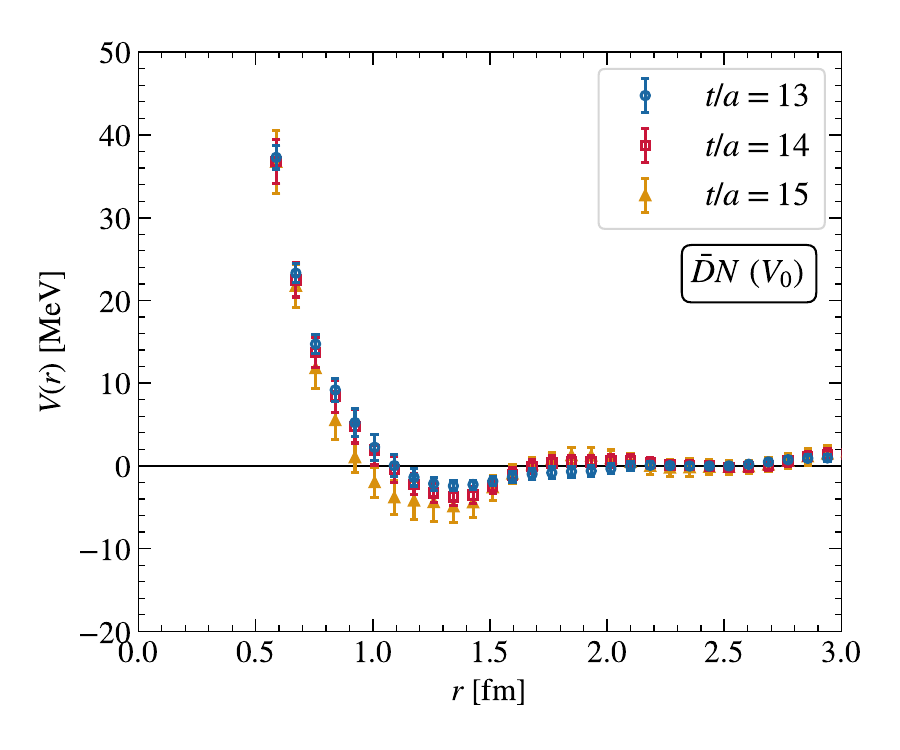}}
    \subfloat{\includegraphics[width=0.50\linewidth]{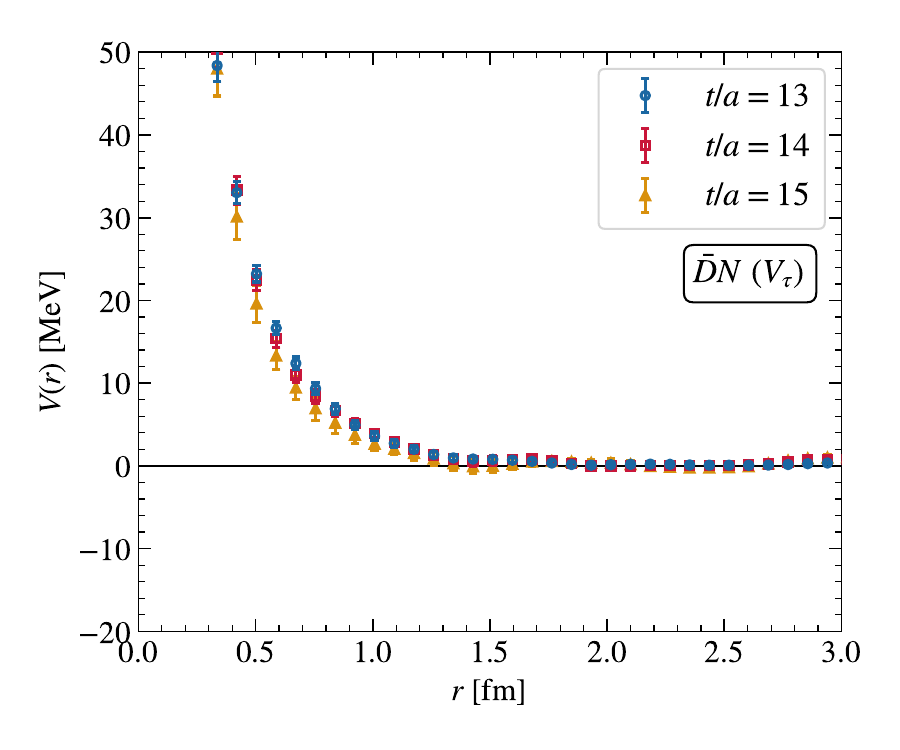}}
    \caption{Isospin independent and dependent components $V_0(r)$ and $V_\tau(r)$ at Euclidean time slices $t/a=13$ (blue, circle), $t/a=14$ (red, square) and $t/a=15$ (gold, triangle).}\label{fig:Lopot_op}
\end{figure*}
In Fig.\,\ref{fig:Lopot_op}, we show $V_0(r)$ and $V_\tau(r)$ at Euclidean time slices $t/a = 13$-$15$.
We observe that $V_0(r)$ has an attractive pocket between $r\sim 0.8$ fm to $r\sim 2.0$ fm and 
$V_\tau (r)$ is positive and confined at short distances, $r\lesssim1$ fm.
In the $I=0$ channel, $V_\tau(r)$ appears with a negative sign and acts as a totally attractive force, and for the $I=1$ channel $V_\tau(r)$ contributes with a positive sign and acts as a purely repulsive force.

\subsection{Comparison to $KN$ Systems}
\begin{figure*}[!htb]
    \centering
    \subfloat{\includegraphics[width=0.50\linewidth]{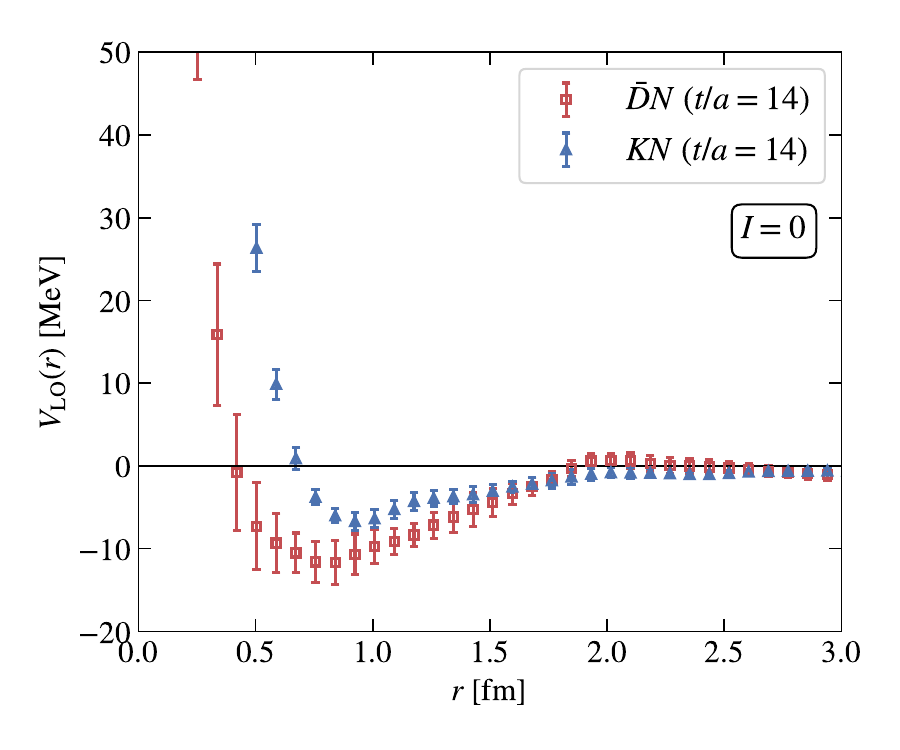}} 
    \subfloat{\includegraphics[width=0.50\linewidth]{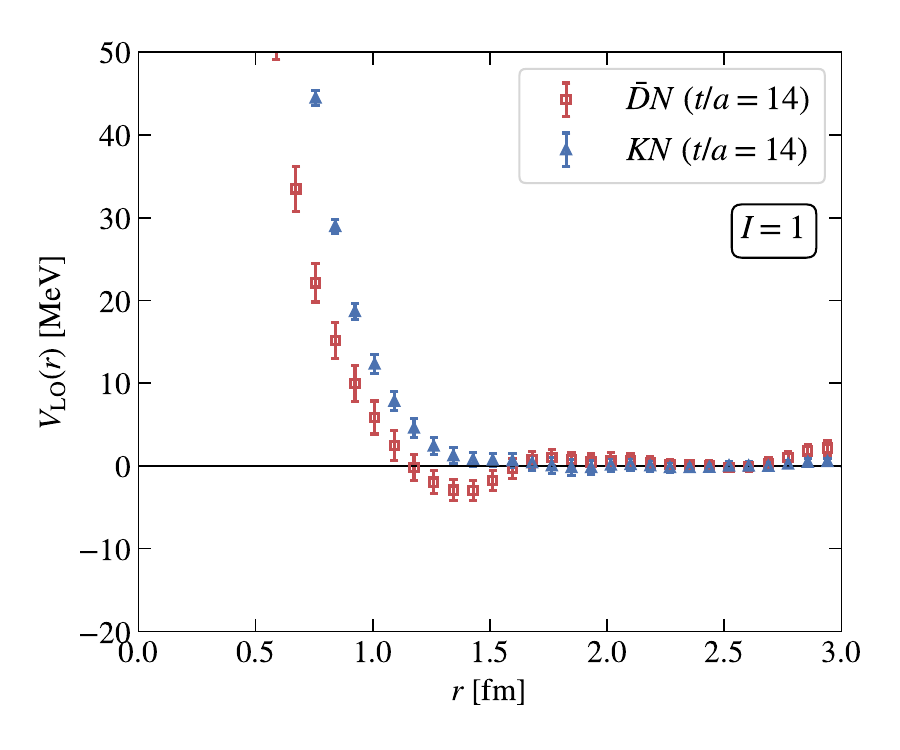}} 
    \caption{$V_{\text{LO}}(r)$ for the $\bar{D}N$ system (red, square) and $KN$ system (blue, triangle) in the isospin $I=0$ channel (left), and the $I=1$ channel (right) at Euclidean time slice $t/a=14$. The $KN$ potential is taken from Ref.\,\cite{Murakami:2025owk}.
    }\label{fig:Lopot_vsKN}
\end{figure*}
The $KN$ system is the strange-sector analogue of $\bar{D}N$, replacing the anti-charm quark with an anti-strange quark. 
In Fig.~\ref{fig:Lopot_vsKN}, we show the LO potentials of $\bar{D}N$ and $KN$ in the $I=0$ and $I=1$ channels at Euclidean time slice $t/a=14$. 
The $KN$ potential, taken from Ref.\,\cite{Murakami:2025owk}, was obtained by the HAL QCD method using the same lattice setup and statistics. 
The qualitative features are the same for other time slices, 
e.g.,
$t/a=13$ or $t/a=15$.

We find that the $\bar{D}N$ system manifests stronger attraction in both isospin channels than the $KN$ system. 
Its repulsive core is confined to a shorter range than that of $KN$. 
In the $I=0$ channel, the attractive pocket is deeper by approximately a factor of two compared to the $KN$ case. 
Moreover, we observe a small attractive pocket in the $I=1$ channel of $\bar{D}N$, whereas no such feature is clearly visible for $KN$.
The difference between the $\bar{D}N$ potential and the $KN$ potential predominantly arises from the heavy-quark dynamics in the $\bar{D}N$ system.
One possible contribution is the $\bar{D}N$-$\bar{D}^{\ast}N$ channel-coupling effect
discussed in phenomenological models \cite{Haidenbauer:2007jq,Yasui:2009bz,Gamermann:2010zz,Yamaguchi:2011xb,Yamaguchi:2022oqz},
while other contributions which are not directly related to this channel-coupling effect may also be relevant.
In order to quantitatively determine the contribution from the $\bar{D}N$-$\bar{D}^{\ast}N$ channel coupling,
an explicit coupled-channel analysis is required,
which we leave for future work.

\section{Scattering Quantities of $\bar{D}N$ System in $s$-wave} \label{sec:scattpar}
In this section we obtain the $s$-wave phase shifts, scattering length and effective range from the potential given in Section \ref{sec:lopot}. 
To extract the $s$-wave phase shifts, we parameterize $V_{\text{LO}}(r)$ using two functional forms. 
A purely phenomenological three- (four-) Gaussian form, and a two- (three-) Gaussian form with an additional two-pion exchange term are used for the $I=0$ ($I=1$) channel. 
In the following, we refer to these fits as the "multi-gauss fit" and the "multi-gauss+TPE fit", respectively.
The multi-gauss+TPE fit is motivated by the fact that two-pion exchange (TPE) is the longest-range mode since one-pion exchange is prohibited by parity conservation.
The same fitting form was used for the TPE fit in the HAL QCD analysis of the $KN$ system \cite{Murakami:2025owk}.

\begin{figure*}[!htb]
    \centering
    \subfloat{\includegraphics[width=0.5\linewidth]{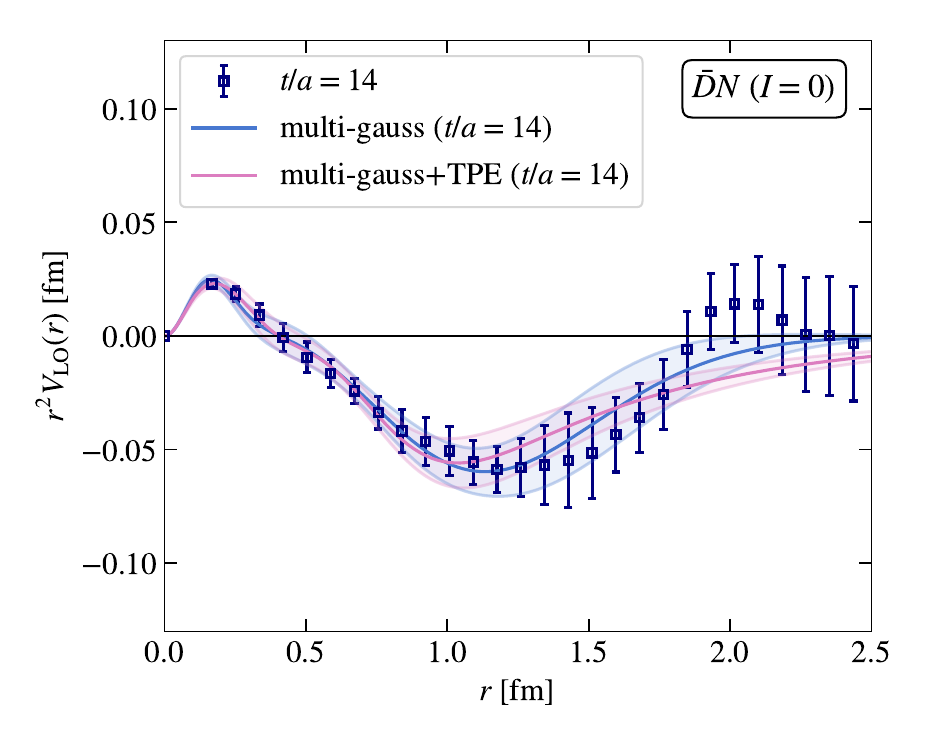}}
    \subfloat{\includegraphics[width=0.5\linewidth]{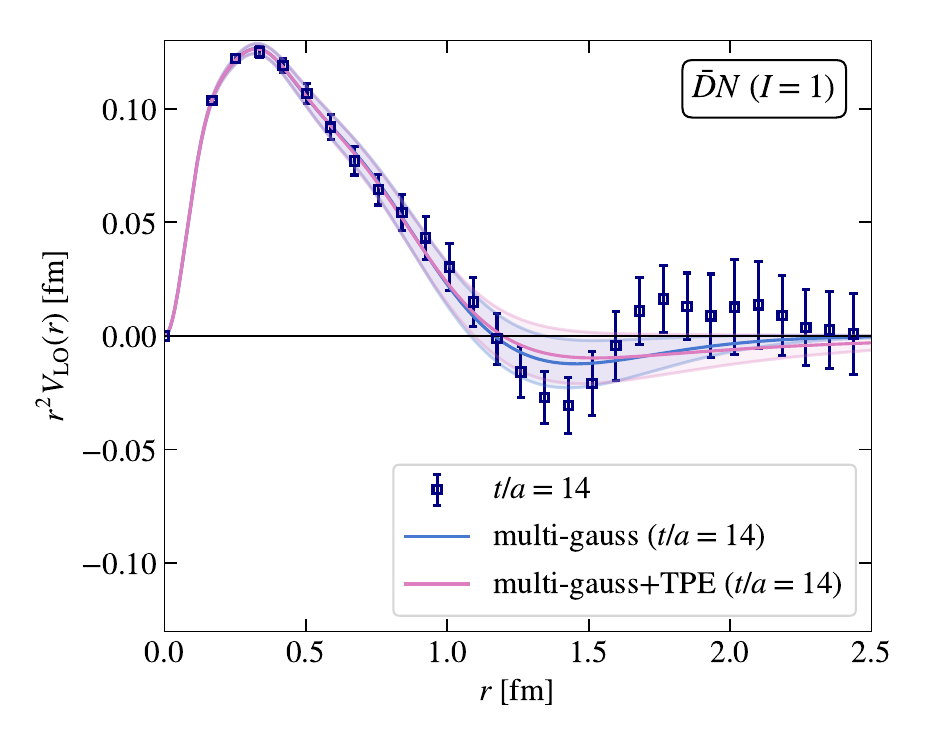}}
    \caption{$r^2 V_{\text{LO}}(r)$ for the $I=0$ (left) and $I=1$ (right) channels at Euclidean time slice $t/a=14$. 
    The blue and pink bands represent the multi-gauss fit and the multi-gauss+TPE fit, respectively.
    }
    \label{fig:fit4gvsfit3gtpe}
\end{figure*}

The parameterizations of the multi-gauss fit and the multi-gauss+TPE fit are explicitly given by
\begin{align}
    V_{\text{fit}}^{\textrm{MG}}(r) &=\sum_{i=1}^{3(4)} a_i~e^{-(r/b_i)^2}\label{eq:4g}\\
    V_{\text{fit}}^{\textrm{MG+TPE}}(r) &= \sum_{i=1}^{2(3)} c_i~e^{-(r/d_i)^2}+c_{\textrm{TPE}}(1-e^{-\Lambda^2r^2})^2\frac{e^{-2m_\pi r}}{r^2},\label{eq:3gTPE}
\end{align}
respectively, 
where $\{a_i\}$, $\{b_i\}$, $\{c_i\}$, $\{d_i\}$ and $c_{\textrm{TPE}}$ are real-valued fitting parameters and $m_\pi=137$ MeV.
$(1-e^{-\Lambda^2r^2})^2$ is a form factor to regularize the TPE term at the origin.
The $\Lambda$ dependence of the resulting phase shifts has been studied at $\Lambda=400$, $658$ and $1000$ MeV, and found to be small relative to the dependence on the fit function. In the following, we fix $\Lambda=658$ MeV ($\Lambda^{-1}=0.3$ fm).
The fitting parameters are determined by performing an uncorrelated fit with a fitting range chosen as the interval from $r=0$ fm to $r= 2.7$ fm.
For $V_{\text{LO}}(0)$, we substitute the values from the $A_1^+$ projected R-correlator, since the Misner method is not applicable at the origin~\cite{Miyamoto:2019jjc}.
The effect of this prescription is negligible when calculating the phase shifts in the low-energy region.
The fitted parameters for the multi-gauss fit and the multi-gauss+TPE fit at the time slice $t/a=14$ are listed in Tabs.\,\ref{tab:fpars_4g} and \ref{tab:fpars_3gtpe} together with their $\chi^2$/dof values, respectively. 
The $\chi^2$/dof values for the multi-gauss+TPE fit are larger than the multi-gauss fit, but are still comparable in magnitude.
In Fig.\,\ref{fig:fit4gvsfit3gtpe}, we show the multi-gauss fit and multi-gauss+TPE fit for $r^2 V_{\text{LO}}(r)$ at
$t/a=14$.
We see that the multi-gauss+TPE fit has a longer attractive tail compared to that of the multi-gauss fit, and thus we expect larger attractive behavior in the phase shifts at low energies.

\begin{table*}[!htb]
    \centering
    \caption{Fitted parameters of the multi-gauss fit for the $I=0$ and $I=1$ channels at time slice $t/a=14$.
    Values in parentheses are the statistical uncertainties.
    }
    \label{tab:fpars_4g}
    \begin{tabular}{cccccccccc} \hline\hline
         $I$&$\chi^2$/dof&$a_1$ [MeV]&$a_2$ [MeV]&$a_3$ [MeV]&$a_4$ [MeV]&$b_1$ [fm]&$b_2$ [fm]&$b_3$ [fm]&$b_4$ [fm] \\ \hline
         0 & 0.45(28) & 490(18) & 107(50) & -97(33) & --- & 0.161(5) & 0.702(78) & 0.860(95) & ---  \\ 
         1 & 0.68(51) & 1163(29) & 557(30) & 154(20) & -24(12) & 0.130(1)&0.275(8)&0.620(40)& 0.902(62)  \\ \hline\hline
    \end{tabular}
\end{table*}
\begin{table*}[!htb]
    \centering
    \caption{Fitted parameters of the multi-gauss+TPE fit for the $I=0$ and $I=1$ channels at time slice $t/a=14$. $m_\pi$ and $\Lambda$ are fixed to 137 MeV and 638 MeV, respectively.
    Values in parentheses are the statistical uncertainties.
    }
    \label{tab:fpars_3gtpe}
    \begin{tabular}{ccccccccc} \hline\hline
         $I$&$\chi^2$/dof&$c_1$ [MeV]&$c_2$ [MeV]&$c_3$ [MeV]&$d_1$ [fm]&$d_2$ [fm]&$d_3$ [fm]&$c_{\textrm{TPE}}$ [MeV fm$^{2}$]\\ \hline
         0 & 0.60(51) & 215(53) & 294(58) & --- & 0.108(23) & 0.473(16) & --- & -58(14) \\ 
         1 & 0.83(56) & 1130(41) & 527(31) & 192(41) & 0.129(1) & 0.289(27) & 0.575(36) & -19(21) \\ \hline\hline
    \end{tabular}
\end{table*}

\begin{figure*}[!htb]
    \centering   
    \subfloat{\includegraphics[width=0.5\linewidth]{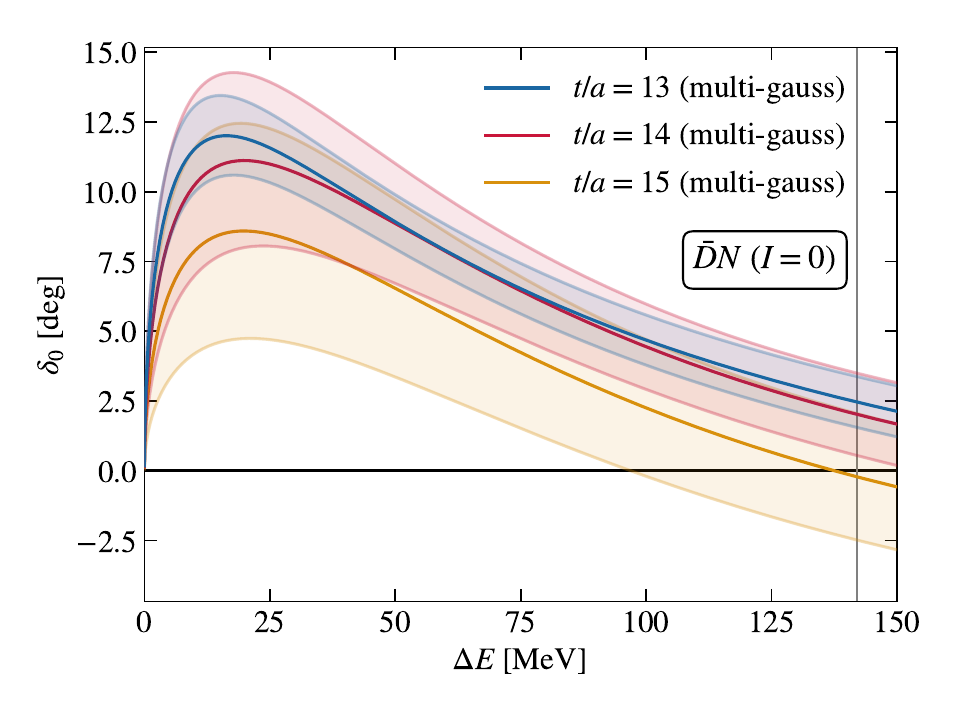}}
    \subfloat{\includegraphics[width=0.5\linewidth]{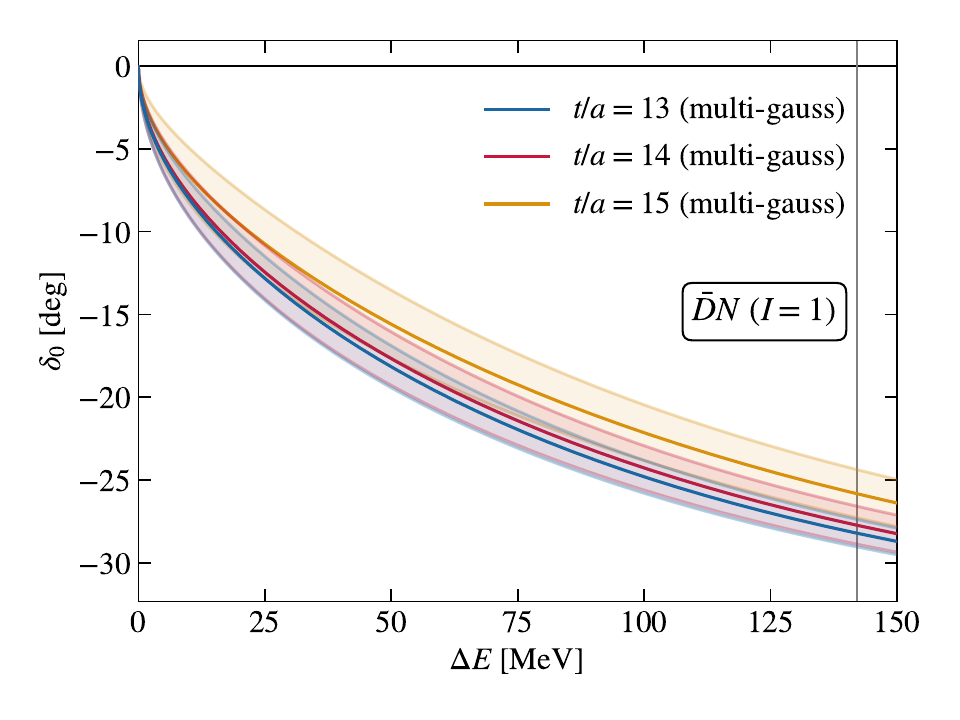}}
    \caption{$s$-wave phase shifts $\delta_0$ 
    as a function of $\Delta E = E-m_{D}-m_{N}$
    for $I=0$ (left) and $I=1$ (right) channels at Euclidean times $t/a=13$ (blue), $t/a=14$ (red) and $t/a=15$ (gold) obtained from the multi-gauss fit.
    The vertical lines at $\Delta E\simeq 140$ MeV 
    show the inelastic threshold of $\bar{D}^\ast N$.
    }\label{fig:phaseshift_4g}
\end{figure*}

\begin{figure*}[!htb]
    \centering
    \subfloat{\includegraphics[width=0.5\linewidth]{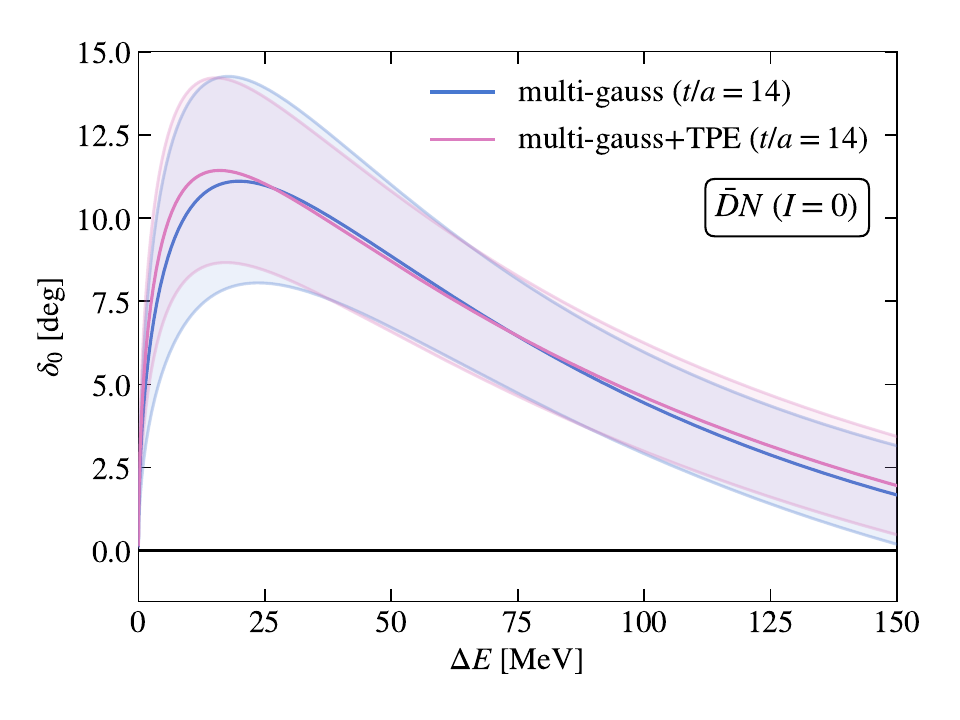}}
    \subfloat{\includegraphics[width=0.5\linewidth]{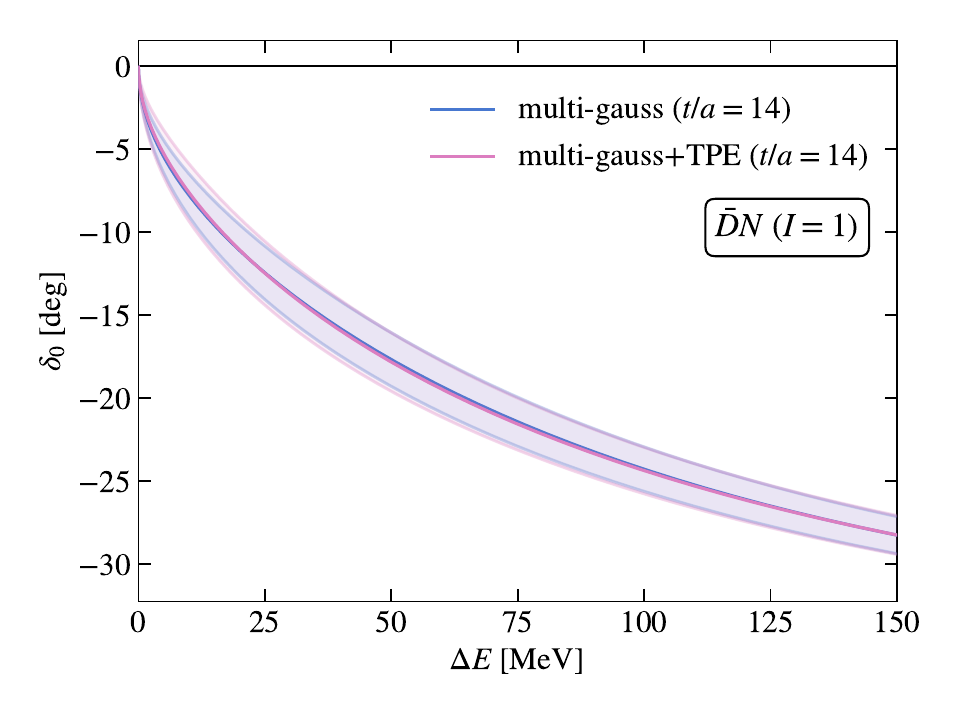}}
    \caption{$s$-wave phase shifts $\delta_0$ 
    as a function of $\Delta E = E-m_{D}-m_{N}$
    for $I=0$ (left) and $I=1$ (right) channels obtained from the multi-gauss fit (blue) and multi-gauss+TPE fit (pink) at the Euclidean time slice $t/a=14$.
    }
    \label{fig:fit4gvsfit3gtpe_delta}
\end{figure*}

We then obtain the $s$-wave phase shifts of $\bar{D}N$ scattering by solving Eq.\,\eqref{eq:Schrodinger} in the infinite volume with $V_{\text{fit}}^{\textrm{MG}}(r)$ and $V_{\text{fit}}^{\textrm{MG+TPE}}(r)$ at $t/a=13$-$15$.
Phase shifts $\delta_0$ obtained from $V_{\text{fit}}^{\textrm{MG}}(r)$ for $I=0$ and $I=1$ channels are shown in Fig.\,\ref{fig:phaseshift_4g}. 
We also show the comparison of the phase shifts from $V_{\text{fit}}^{\textrm{MG}}(r)$ and $V_{\text{fit}}^{\textrm{MG+TPE}}(r)$ at $t/a=14$ in Fig.\,\ref{fig:fit4gvsfit3gtpe_delta}.
The time-slice dependence of $\delta_0$ is found to be within the statistical uncertainties, and the fit-function dependence is found to be even smaller.
We see that $\delta_0$ in $I=0$ shows an attractive behavior in the low-energy region, while $\delta_0$ in $I=1$ shows a repulsive behavior in all energy regions. 
The phase shifts obtained from the fitted potentials were numerically confirmed to asymptotically approach zero at sufficiently large energies, indicating $\Delta\delta_0=\delta_0(0)-\delta_0(\infty)=0$.
Thus from Levinson's theorem we conclude there are no bound states in both isospin channels.
We also do not observe any rapid rise in the phase shift crossing $\pi/2$, indicating 
the absence of narrow resonances.
Furthermore, we searched for poles of the scattering amplitude in the complex energy plane over the region corresponding to 
$-2 \leq {\rm Re} (k/m_\pi), {\rm Im} (k/m_\pi) \leq 2$, 
and confirmed the absence of near-threshold resonances.

\begin{figure*}[!htb]
    \centering
    \subfloat{\includegraphics[width=0.5\linewidth]{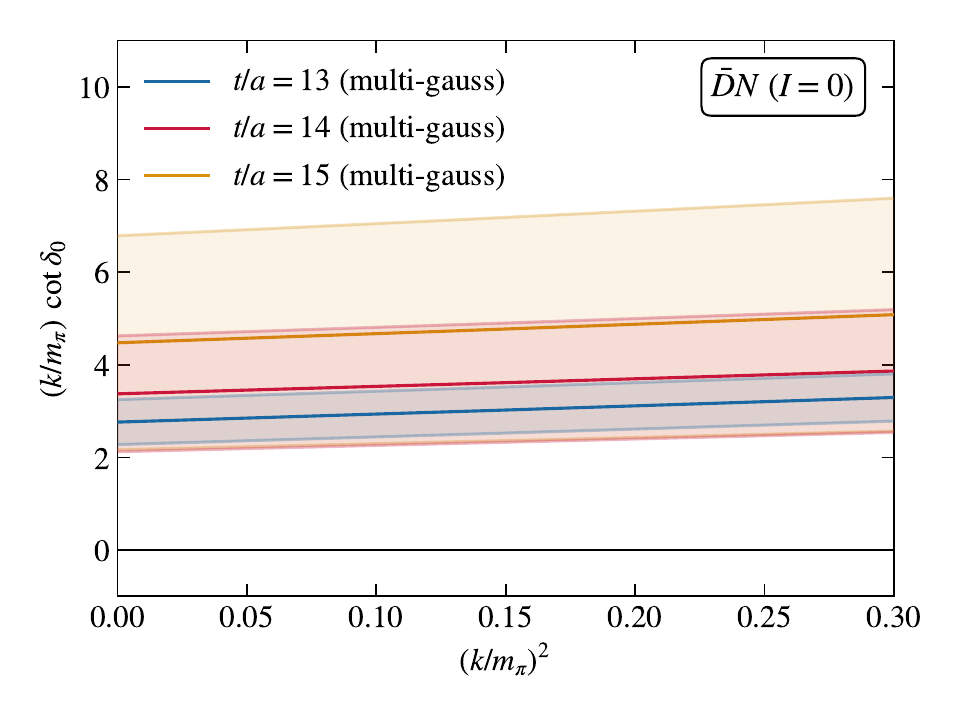}}
    \subfloat{\includegraphics[width=0.5\linewidth]{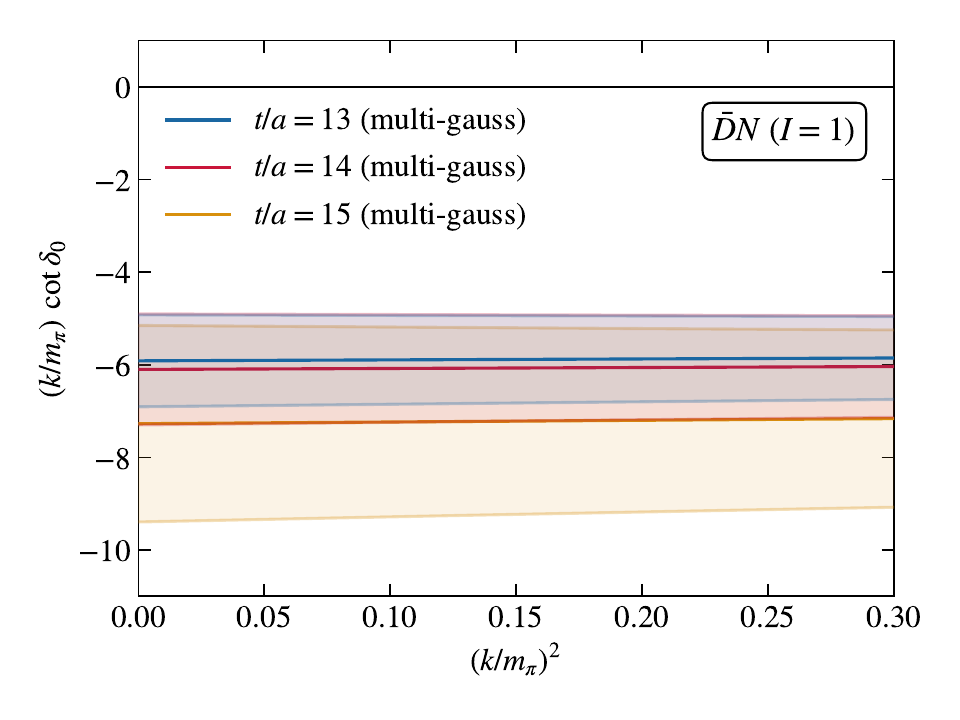}}
    \caption{$k\cot\delta_0/m_\pi$ in terms of $k^2/m_\pi^2$ for $I=0$ (left), and $I=1$ (right) at Euclidean times $t/a=13$ (blue), $t/a=14$ (red) and $t/a=15$ (gold) from the multi-gauss fit.
    }\label{fig:kcot_4g}
\end{figure*}
When $k$ is sufficiently small, $k\cot\delta_0$ can be expanded by the effective range expansion (ERE) as
\begin{gather}
    k\cot\delta_0=\frac{1}{a_0}+\frac{1}{2}r_{\text{eff}}k^2+\mathcal{O}(k^4), \label{eq:ERE}
\end{gather}
where $a_0$ ($r_{\text{eff}}$) is the scattering length (effective range).
$k\cot\delta_0$ obtained from $V_{\text{fit}}^{\textrm{MG}}(r)$ for the $I=0$ ($I=1$) channel are shown in Fig.\,\ref{fig:kcot_4g} 
where $1/a_0$ corresponds to the intercept to the vertical axis, and $r_{\text{eff}}$ corresponds to the slope of the line, respectively. 
The scattering length and effective range are determined by fitting $k\cot\delta_0$ discarding the higher-order terms of $\mathcal{O}(k^4)$ in Eq.\,\eqref{eq:ERE}.   
Varying the fitting range from $(k/m_\pi)^2\in[0.00,0.10]$ to $(k/m_\pi)^2\in[0.00,0.30]$ (corresponding to $\Delta E\in[0.00,1.50]$ MeV to $\Delta E\in[0.00,4.49]$ MeV), which is well below the $\bar{D}^\ast N$ threshold, we confirm that the fitting range uncertainties in the scattering length and effective range are small for both the multi-gauss fit case and the multi-gauss+TPE fit cases. 
These observations indicate that the ERE truncated at the $k^2$ term provides a reasonable description of $k\cot\delta_0$ in this kinematic region.

Using the values at $t/a=14$ with a fitting range of $(k/m_\pi)^2\in[0.00,0.20]$ as centre values, we obtain,
$a_0^{I=0}= 0.425 \pm 0.163 (_{-0.105}^{+0.094})~\text{fm}$, $r_{\text{eff}}^{I=0}= 4.63 \pm 0.79 (_{-0.05}^{+1.06})~\text{fm}$, $a_0^{I=1}= -0.235 \pm 0.046 (_{-0.007}^{+0.038})~\text{fm}$ and $r_{\text{eff}}^{I=1}= 0.61 \pm 1.05 (_{-0.02}^{+0.47})~\text{fm}$ for the multi-gauss fit case, 
and 
$a_0^{I=0}= 0.513 \pm 0.151~(_{-0.066}^{+0.012})~\text{fm}$, $r_{\text{eff}}^{I=0}= 5.69 \pm 1.08~(_{-0.21}^{+1.06})~\text{fm}$, $a_0^{I=1}= -0.222 \pm 0.071~(_{-0.006}^{+0.035})~\text{fm}$ and $r_{\text{eff}}^{I=1}= 1.99 \pm 3.75 ~(_{-0.11}^{+1.14})~\text{fm}$ for the multi-gauss+TPE fit case, 
where the second and third terms are the statistical and systematic errors, respectively.
The systematic errors are evaluated by considering the time-slice dependence between $t/a=13$, $15$ and $t/a=14$, together with the ERE fit-range dependence in $k^2$.
$a_0^{I=0}$ ($a_0^{I=1}$) are positive (negative) manifesting the attractive (repulsive) behavior of the $\bar{D}N$ interaction in the low-energy region.

We consider other sources of systematic errors,
(i) finite volume effect of the periodic box, (ii) the interpolation to physical charm quark mass, (iii) truncation of the derivative expansion of the interaction kernel. 
First,
the finite volume effect is expected to be negligible because of the large volume of the box ($L/2=4.06$ fm) compared to the interaction range ($\sim 2.0$ fm).
We note that the large effective range extracted from the phase shift for $I=0$ does not represent the spacial extent of the interaction, and therefore does not imply large finite-volume effects.
Second,
the relative difference between the LO potentials obtained from the
two charm quark masses (Set-I and Set-II) is 
the order of one percent, and is much smaller relative to the statistical uncertainties of the potential.
Therefore, a linear interpolation should be an appropriate interpolation to the physical charm quark setup.
The interpolation to physical charm quark mass can be performed at different steps, for example, at the level of the R-correlators, the potentials, or the phase shifts. 
The associated systematic uncertainty from interpolating at different steps, however, has been confirmed to be very small ($\ll 1\%$).
Third,
in order to study the truncation error of the derivative expansion, we 
evaluated the finite-volume energy levels from our LO potential, and compared them to the effective energy calculated directly from the R-correlators as shown in Appendix\,\ref{sec:fvenergy}. 
The ground state energy calculated from the LO potential agrees with the effective energy
directly obtained from the R-correlators, implying that the LO potential gives a reasonable description in the low-energy region relevant to the ground state of the $\bar{D}N$ system.
A more quantitative assessment of the contributions from higher-order terms in the derivative expansion, particularly in the higher-energy region, would require, for example, employing correlators computed with different source operators as practiced in previous HAL QCD studies~\cite{HALQCD:2018gyl}, and is beyond the scope of this study.
In addition, assessment of possible nonlocal contributions to the interaction kernel at energies near the $\bar{D}^\ast N$ threshold would require a $\bar{D} N$-$\bar{D}^\ast N$ coupled-channel analysis, which should be pursued in future work.

In Tab.\,\ref{tab:scatt_pars_fin}, we present our final results on the $\bar{D}N$ scattering length and effective range.
For the centre values and statistical errors we opt for the results from the multi-gauss fit, since the results have smaller $\chi^2/\text{dof}$ values.
The systematic error is determined by 
combining in quadrature
the systematic error of the multi-gauss fit and the difference between the multi-gauss fit and multi-gauss+TPE fit.

\begin{table}[!htb]
    \caption{Scattering length and effective range of the $\bar{D}N$ system in $s$-wave. The second (third) term is the statistical (systematic) uncertainties.}
    \label{tab:scatt_pars_fin}
    \begin{tabular}{ccc} \hline\hline
         $I$ & $a_0$ [fm] & $r_{\text{eff}}$ [fm] \\ \hline
         $0$ & $ 0.425 \pm 0.163 (_{-0.105}^{+0.129})$ & $4.63 \pm 0.79 (_{-0.05}^{+1.50})$\\ 
         $1$ & $-0.235 \pm 0.046 (_{-0.007}^{+0.040})$ & $0.61 \pm 1.05 (_{-0.02}^{+1.46})$\\ \hline\hline
    \end{tabular}
\end{table}

\subsection{Comparison of Scattering Length to Model Studies}
\begin{figure}
    \centering
    \includegraphics[width=0.8\linewidth]{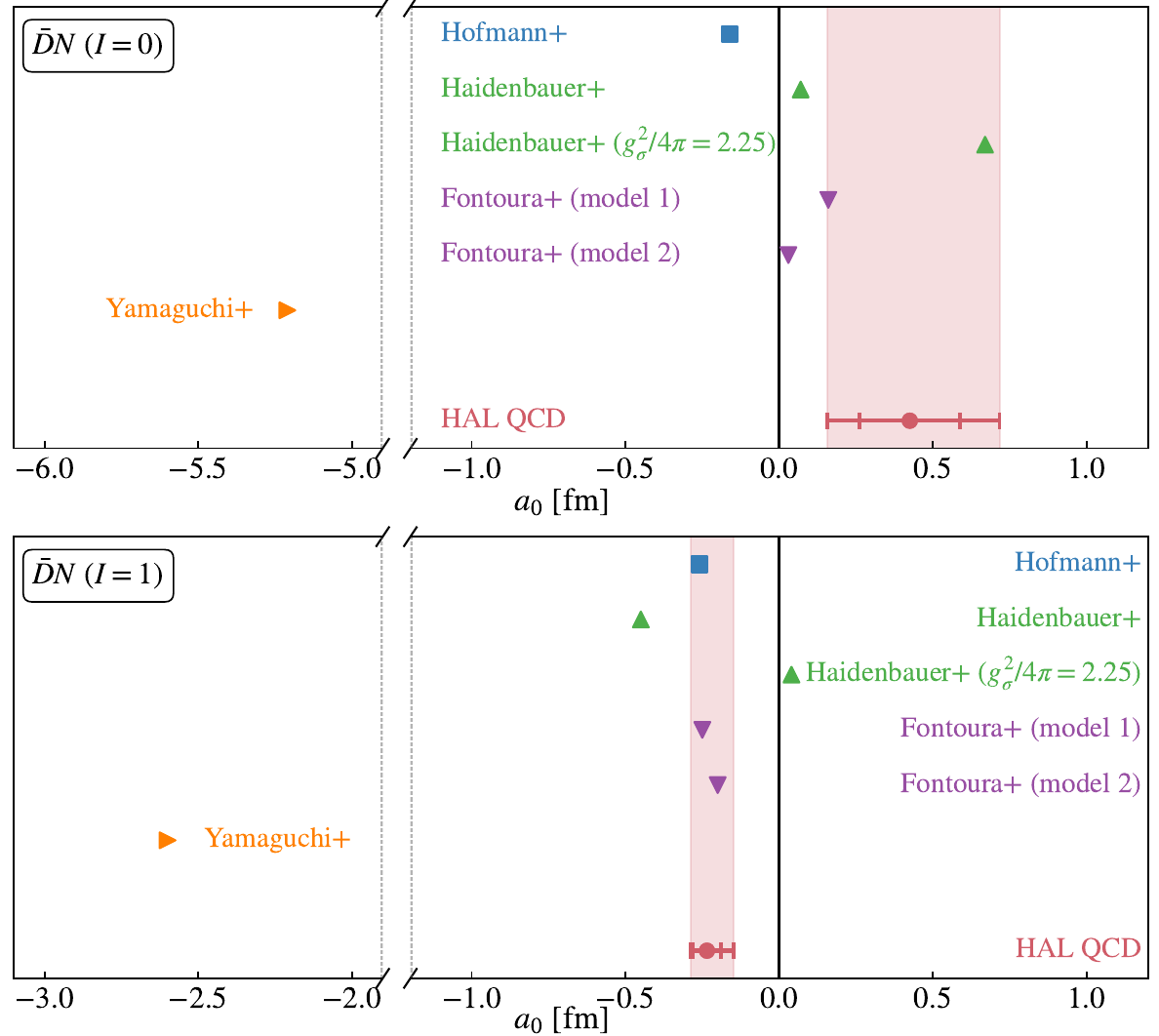}
    \caption{Comparison of $\bar{D}N$ scattering length to model studies \cite{Hofmann:2005sw,Haidenbauer:2007jq,Fontoura:2012mz,Yamaguchi:2022oqz}.
    The red point (and band) is our results given in Tab.\,\ref{tab:scatt_pars_fin}. 
    The inner error bars represent the statistical uncertainties and the outer error bars represent the statistical and systematic uncertainties combined in quadrature.
    "Haidenbauer+ $(g_\sigma^2/4\pi=2.25)$" is the result from the model Ref.\,\cite{Haidenbauer:2007jq} with increased $g_\sigma^2$ where the value is given in Ref.\,\cite{ALICE:2022enj}.
    }
    \label{fig:compareEFT}
\end{figure}

In Fig.\,\ref{fig:compareEFT}, we show a comparison of the $\bar{D}N$ scattering lengths obtained in the present study with results from several models where the scattering lengths are available in literature. 
The results of the following models are presented.
A model with SU(4) contact interaction and vector-meson exchange \cite{Hofmann:2005sw}, meson-exchange models with an explicit short-distance quark-gluon exchange interaction~\cite{Haidenbauer:2007jq, Fontoura:2012mz} and a meson-exchange model respecting both chiral and heavy-quark spin symmetry~\cite{Yamaguchi:2022oqz}.
We observe that the scattering lengths from our study are within the same order of magnitude to the model studies except for Ref.\,\cite{Yamaguchi:2022oqz} which exhibits a bound state.
Our results can be used to constrain model parameters and further improve theoretical models concerning the $\bar{D}N$ interaction.
%

\section{Summary and Outlook}\label{sec:conclusion}
We study the $\bar{D}N$ scattering from lattice QCD simulations for the first time with gauge configurations generated at the physical point with a pion mass $m_\pi\simeq 137$ MeV.
Using the HAL QCD method, we obtain the leading-order $\bar{D}N$ potential in the derivative expansion for the isospin $I=0$ and $I=1$ channels in $s$-wave.
The $\bar{D}N$ potentials in both 
isospin channels manifest a repulsive core at short distances with an attractive pocket at intermediate to long distances, while the $I=0$ channel shows stronger attraction than the $I=1$ channel.
We also compare the $\bar{D}N$ potentials to the $KN$ potentials in Ref.\,\cite{Murakami:2025owk} obtained using the same configurations, and find that $\bar{D}N$ shows stronger attraction than $KN$ in both isospin channels. 
We also obtain the $s$-wave phase shifts, scattering length and effective range 
from the $\bar{D}N$ potentials.
The $s$-wave phase shift of the $I=0$ channel shows weak attraction in the low-energy region with a positive scattering length of $a_0= 0.425 \pm 0.163 (_{-0.105}^{+0.129})$ fm, whereas the $I=1$ channel shows repulsion in all energy regions with a negative scattering length of $a_0=-0.235 \pm 0.046 (_{-0.007}^{+0.040})$ fm. 
Our study finds that the attractions between the $\bar{D}$ meson and the nucleon are small, and therefore no $s$-wave bound states (pentaquark states) exist in the $I=0$ or $I=1$ channels of the $\bar{D}N$ system.

The present study is performed within a single-channel framework using the potential at leading order in the derivative expansion.
As future directions, it is important to extend the present analysis to the coupled $\bar{D}N$-$\bar{D}^{\ast}N$ system to quantitatively determine the coupled-channel effects. It is also important to investigate higher-order terms in the derivative expansion to further examine the nonlocality of the interaction.
The coupled-channel analysis can also be extended to the $BN$-$B^\ast N$ system.
Especially in the $BN$-$B^\ast N$ channel, we expect that the channel-coupling effects will play a more important role in yielding stronger attraction \cite{Yamaguchi:2022oqz}. 
Another interesting study would be the $s$-wave scattering of $\bar{D}^\ast N$ in the $I(J^P) = 0(3/2^-)$ channel, as a theoretical study \cite{Yamaguchi:2011xb} predicts strong attraction leading to a Feshbach resonance decaying to $\bar{D}N$ in the $d$-wave. 
In addition, applications of our $\bar{D}N$ potential to multi-nucleon systems with $\bar{D}$ mesons, e.g., $\bar{D}NN$ systems, $\bar{D}$-meson nuclei systems, would be beneficial in studying possible charmed nuclei.
Finally, we plan to apply our $\bar{D}N$ potential to analyze the $D^- p$ momentum correlation function in nucleus-nucleus collisions,
for which precise experimental data are expected  
from LHC RUN-3 by the ALICE collaboration.

\section*{Acknowledgements}
The authors would like to thank the members of the HAL QCD collaboration for the enlightening discussions. 
Wren Yamada is grateful for the stimulating research environment at RIKEN iTHEMS.
The numerical lattice QCD simulations have been performed on 
the supercomputer Fugaku. 
We thank ILDG/JLDG \cite{ldg}, which serves as essential infrastructure in this study.
This work was partially supported by
RIKEN Incentive Research Project (``Unveiling pion-exchange interactions between hadrons from first-principles lattice QCD"), 
Adopting Sustainable Partnerships for Innovative Research Ecosystem (ASPIRE), Grant No. JPMJAP2318,
HPCI System Research Project
(hp200130, hp210165, hp220174, hp220066, hp230207, hp230075, hp240213, hp240157,
hp250224, hp250195),
the JSPS
(Grants No. JP22H00129, JP22H04917, JP23H05439, JP25K17384)
``Priority Issue on Post-K computer'' (Elucidation of the Fundamental Laws and Evolution of the Universe), 
``Program for Promoting Researches on the Supercomputer Fugaku'' (Simulation for basic science: from fundamental laws of particles to creation of nuclei) 
and (Simulation for basic science: approaching the new quantum era) (Grants No. JPMXP1020200105, JPMXP1020230411), 
and Joint Institute for Computational Fundamental Science (JICFuS).


\appendix

\section{Effective mass of the $D$ ($D^\ast$) meson}\label{sec:dmeson_mass}
\begin{figure}[!htb]
    \centering
    \subfloat{\includegraphics[width=0.5\linewidth]{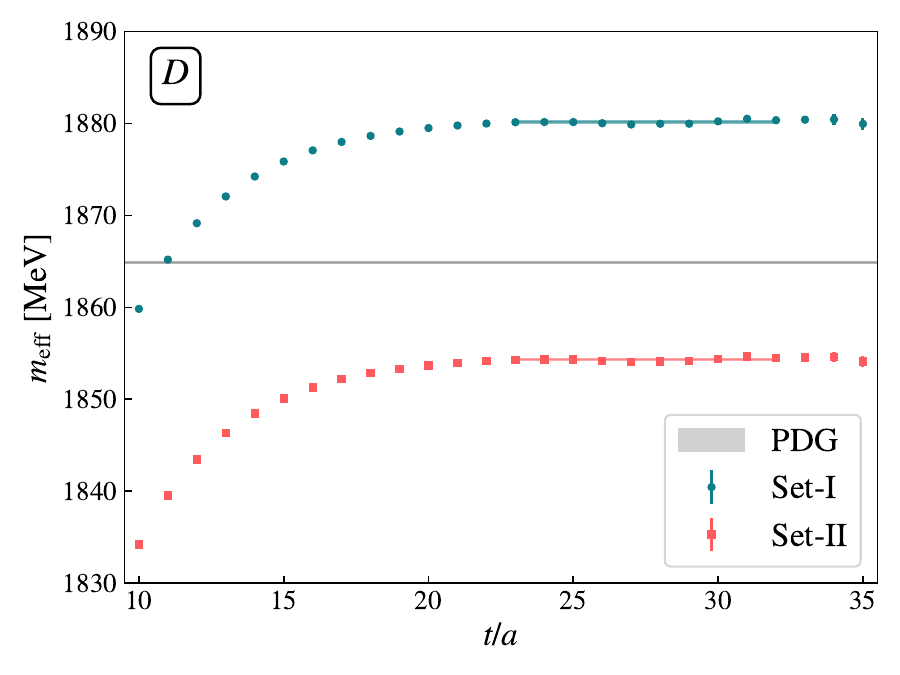}}
    \subfloat{\includegraphics[width=0.5\linewidth]{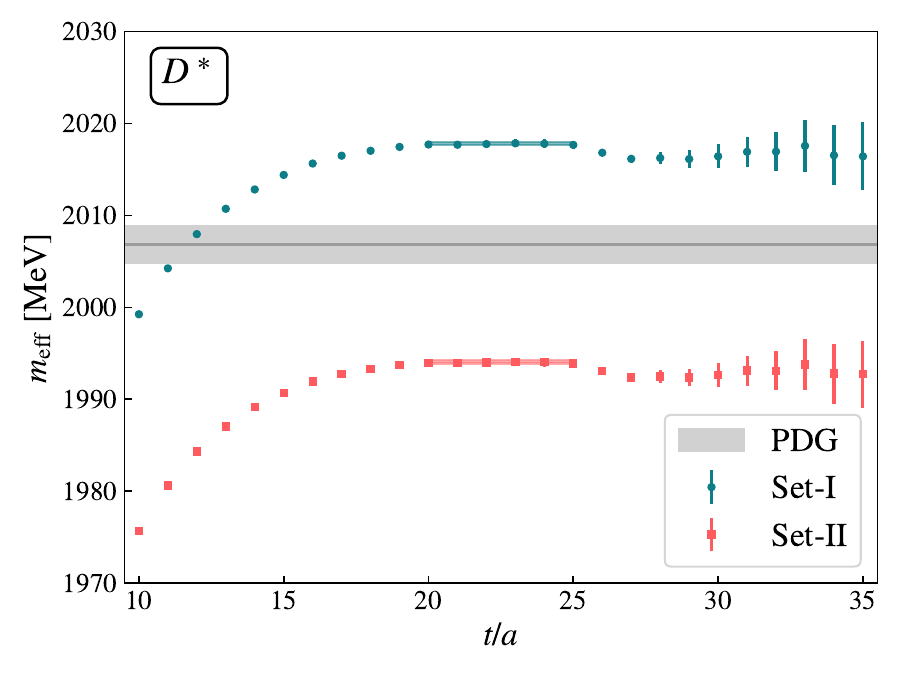}}
    \caption{Effective masses of the $\bar{D}$ meson (left) and the $\bar{D}^\ast$ meson (right) from the two-point correlation functions of Set-I and Set-II. The gray band corresponds to the PDG values \cite{ParticleDataGroup:2024cfk}.}\label{fig:dbar_mass}
\end{figure}

The effective masses of the $\bar{D}$ ($\bar{D}^\ast$) meson in Set-I and Set-II are shown in Fig.\,\ref{fig:dbar_mass}.
We determine their masses $m_D$ ($m_{D^\ast}$) by fitting their two-point correlation functions using the single cosh function on the time interval of $t/a\in[23,32]$ ($t/a\in[20,25]$), which are given in Table\,\ref{tab:meson_mass}. 
As seen, the $\bar{D}$ ($\bar{D}^\ast$) meson mass in Set-I is larger than the experimental (PDG) value~\cite{ParticleDataGroup:2024cfk}, while the $\bar{D}$ ($\bar{D}^\ast$) meson mass in Set-II is smaller. 
It must be noted that the $\bar{D}^\ast$ is not a stable particle at the physical pion mass and its mass should ideally be obtained from the pole position of the $\bar{D}\pi$ scattering matrix in $p$-wave. 
However the energy of lowest lying $\bar{D}\pi$ scattering state on the lattice, approximated by free-particle scattering, is $\sqrt{m_D^2+(2\pi/L)^2}+\sqrt{m_\pi^2+(2\pi/L)^2}=2092~(2066)$ MeV for Set-I (Set-II), which lies considerably above the $\bar{D}^\ast$ mass, so that the effect of the decay is expected to be small.

\begin{table}[!htb]
    \centering
    \caption{Masses of the $\bar{D}$ meson and $\bar{D}^\ast$ meson from Set-I and Set-II in physical units [MeV]. The values in parentheses are the statistical uncertainties. Label ``Experiment'' denotes the PDG values~\cite{ParticleDataGroup:2024cfk} for the charge neutral $D$ and $D^\ast$ mesons.
    }
    \vspace{5pt}\par
    \begin{tabular}{cccc}\hline\hline
          & Set-I [MeV] & Set-II [MeV] & Experiment [MeV] \\ \hline
         $m_{D}$      & 1880.2 (1) & 1854.3 (1) &  1864.84 (5)\\
         $m_{D^\ast}$ & 2017.8 (2) & 1994.1 (2) &  2006.85 (5) \\ \hline\hline
    \end{tabular}
    \label{tab:meson_mass}
\end{table}

\section{Charm-quark mass dependence of the $\bar{D}N$ potential: comparison between Set-I and Set-II}\label{sec:interpolation}
As discussed in Sec.\,\ref{sec:method}, we used two setups for the parameters in the charm quark action. One with a slightly heavier charm quark mass (Set-I) and another with a slightly lighter charm quark mass (Set-II) compared to the physical value.
Here, we compare the potentials for Set-I and Set-II, computed from their respective R-correlators using Eq.\,\eqref{eq:LOpot_}.

\begin{figure*}[!htb]
\centering
\subfloat{\includegraphics[width=0.5\linewidth]{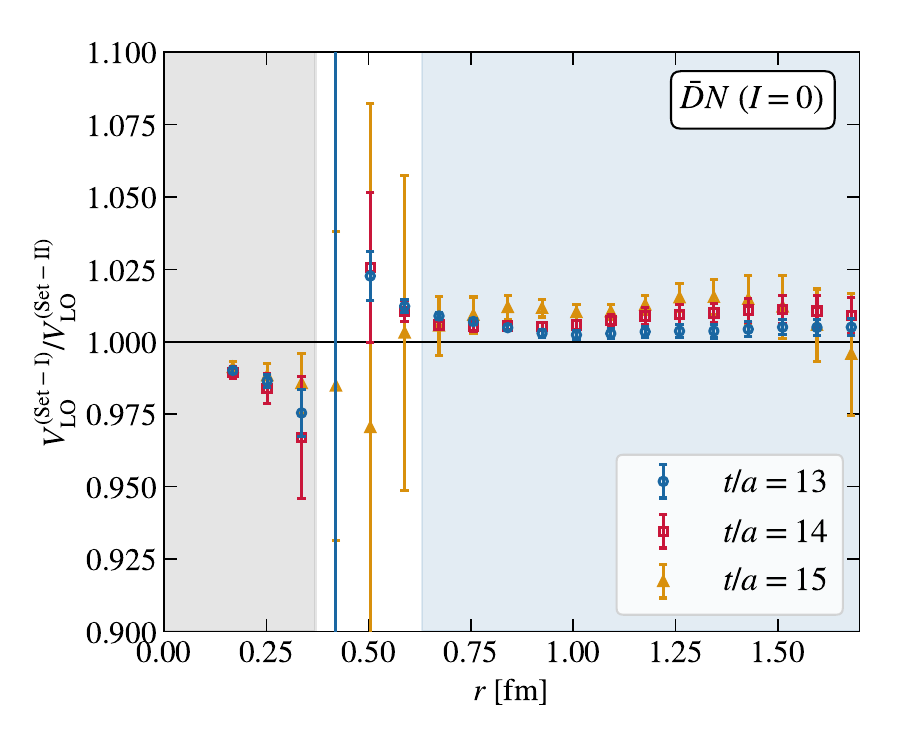}}
\subfloat{\includegraphics[width=0.5\linewidth]{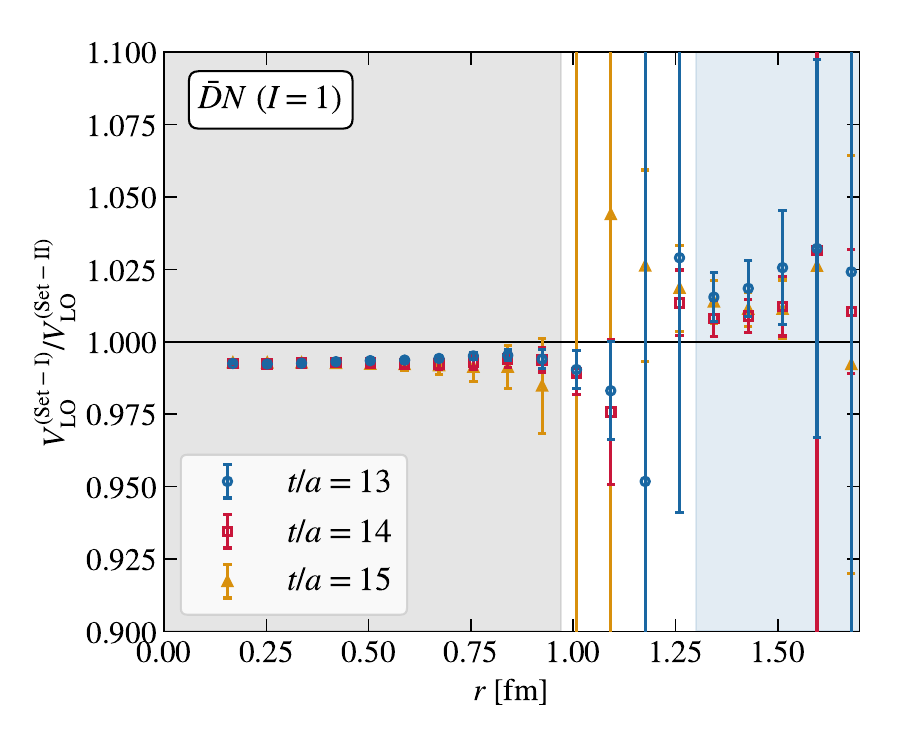}}
\caption{Ratio of the potential $V_{\textrm{LO}}^{\textrm{(Set-I)}}/V_{\textrm{LO}}^{\textrm{(Set-II)}}$ obtained from Set-I and Set-II at time slices $t/a=13$ (blue, circle), $t/a=14$ (red, square), and $t/a=15$ (gold, triangle) for $I=0$ (left) and $I=1$ (right). 
The gray-(blue-) colored domains correspond the domains where $V_{\textrm{LO}}>0$ ($V_{\textrm{LO}}<0$).
}
\label{fig:ratio}
\end{figure*}

Fig.\,\ref{fig:ratio} shows the ratio between $V_{\textrm{LO}}^{\textrm{(Set-I)}}$ and $V_{\textrm{LO}}^{\textrm{(Set-II)}}$ for the time slices $t/a=13$-$15$.  
The divergent behavior around 0.5 fm for the $I=0$ channel and 1.1 fm for the $I=1$ channel arises from the crossover from the repulsive part of the potential ($V_{\textrm{LO}}>0$) and the attractive part ($V_{\textrm{LO}}<0$). 
As we observe, the ratio is smaller than $1$ in the repulsive region while larger in the attractive region, showing that the potential for Set-I has a smaller repulsive core and a deeper attractive pocket compared to that of Set-II.
This behavior indicates that a heavier charm quark mass yields stronger attraction near the physical charm quark mass.
One possible interpretation is that the larger heavy-quark mass leads to a smaller mass gap between $\bar{D}$ and $\bar{D}^{\ast}$ (see Tab.~IV), which may enhance the $\bar{D}N$-$\bar{D}^{\ast}N$ channel-coupling effect. 

The relative difference between $V_{\textrm{LO}}^{\textrm{(Set-I)}}$ and $V_{\textrm{LO}}^{\textrm{(Set-II)}}$ is rather insensitive to $r$ and is approximately $1\%$ which is comparable to the relative mass difference of the $\bar{D}$ meson between Set-I and Set-II ($\sim 1\%$).

\section{Finite-volume energy levels from the HAL QCD potential and the temporal correlators}\label{sec:fvenergy}
To check the systematics from the truncation in the derivative expansion, Eq.\,\eqref{eq:derivexp}, we calculate the ground state energy from the leading-order HAL QCD potential in the finite three-dimensional volume, and compare it to the effective energy obtained directly from the temporal correlators.
Similar analysis for different systems have been performed in prior studies such as Refs.\,\cite{Iritani:2018vfn,Lyu:2022tsd,Murakami:2025owk}.

First we evaluate the ground state energy from the HAL QCD potential in a finite volume.
This can be done by solving the Schr\"{o}dinger equation
\begin{gather}
    H\psi_n(\vec{r})=\frac{k_{n}^2}{2\mu}\psi_n(\vec{r}),\label{eq:box}\\
    H=-\frac{\nabla^2}{2\mu}+V(r),\nonumber
\end{gather}
in a three-dimensional box under a periodic boundary condition.
Here we take $V(r)$ to be $V_{\text{LO}}(r)$ at the time slice $t/a=14$.
The eigen energies $\Delta E_n$ in the $A_1^+$ representation with a box size $L^3=96^3$ are given in Tab.\,\ref{tab:fvenergy}.

\begin{table}[!htb]
    \centering
    \caption{Finite volume eigen energies $\Delta E_n$ ($n=0,1,2$) in the $A_1^+$ representation with a box size $L^3=96^3$ calculated from $V_{\text{LO}}(r)$ at the time slice $t/a=14$.
    }
    \label{tab:fvenergy}
    \begin{tabular}{cccc}\hline\hline
         & $\Delta E_0$ [MeV] & $\Delta E_1$ [MeV] & $\Delta E_2$ [MeV] \\ \hline
         $I=0$ & $-0.141(315)$ & $17.7(5)$ & $36.0(7)$\\
         $I=1$ & $0.304(409)$ & $20.0(4)$ & $39.5(4)$\\\hline\hline
    \end{tabular}
\end{table}

Next we evaluate the effective energy $E_0^{\textrm{eff}}(t)$ from the temporal correlation functions.
We prepare two temporal correlation functions,
\begin{gather}
    R(t)=\sum_{\vec{x}}~R(\vec{x},t),\\
    R^{\textrm{opt}}(t)=\sum_{\vec{x}}~\psi_0^\dagger(\vec{x})R(\vec{x},t),
\end{gather}
where $R(\vec{x},t)$ is the R-correlator defined in Eq.\,\eqref{eq:rcorr}, and $\psi_0$ is the ground state solution of Eq.\,\eqref{eq:box}.
$R^{\textrm{opt}}$ is projected by the ground state of the HAL QCD potential, and is expected to be optimized so that it has a dominant overlap to the actual ground state.
Here we use the linearly-interpolated R-correlator from Set-I and Set-II defined as 
\begin{gather}
    R(\vec{x},t)=w_1 R^{(\textrm{Set-I})}(\vec{x},t)+w_2 R^{(\textrm{Set-II})}(\vec{x},t),
\end{gather}
with the same weights used to interpolate the potential in Sec.\,\ref{sec:lopot}, i.e., $w_1=0.411$ and $w_2=0.589$.
The effective energy is defined as
\begin{gather}
    \Delta E_0^{\textrm{eff}}(t)=\frac{1}{a}\log\biggl[\frac{R^{(\textrm{opt})}(t)}{R^{(\textrm{opt})}(t+1)}\biggr].
\end{gather}
For sufficiently large $t$, $\Delta E_0^{\textrm{eff}}(t)$ should have a plateau at energy $\Delta E_0$ which is the ground state energy measured from the $\bar{D}N$ threshold.

\begin{figure*}[!htb] 
    \centering
    \subfloat{\includegraphics[width=0.5\linewidth]{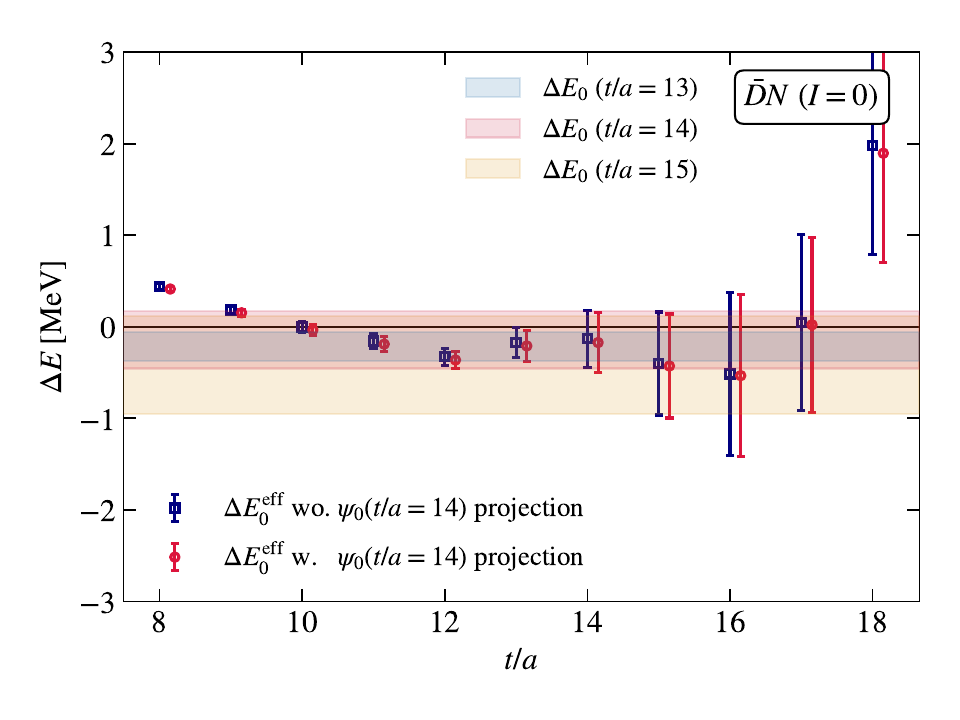}}
    \subfloat{\includegraphics[width=0.5\linewidth]{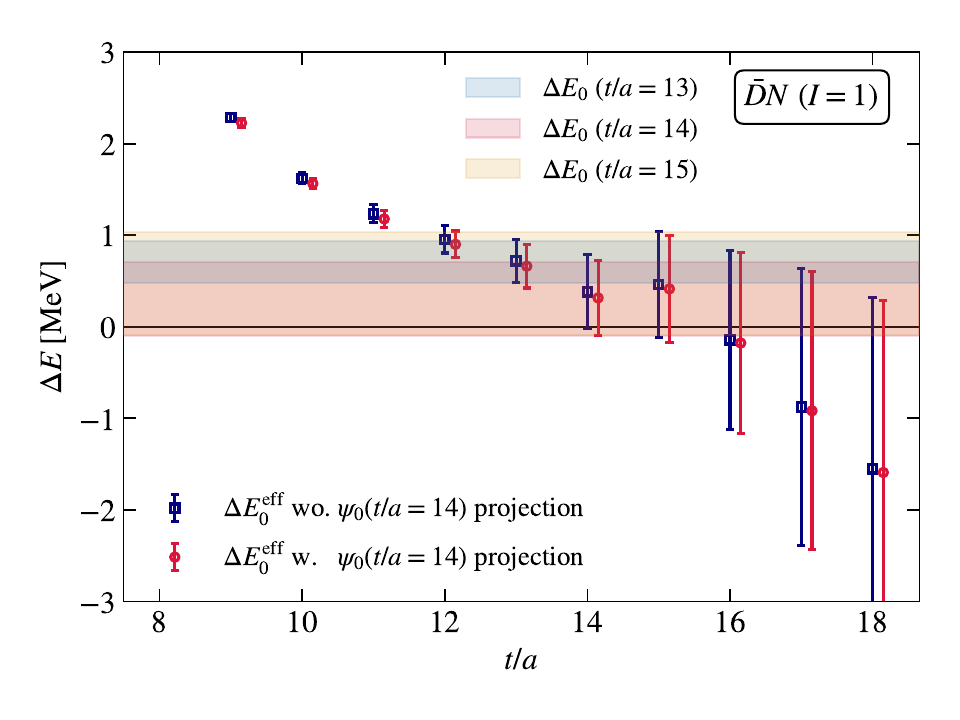}}
    \caption{Effective energies $\Delta E_0^{\text{eff}}$ from correlators $R(t)$ (blue, square) and $R^{\textrm{opt}}(t)$ (red, circle), and the ground state energies $\Delta E_0$ from $V_{\text{LO}}(r)$ at $t/a=13$ (blue band), $14$ (red band), $15$ (gold band) for channels $I=0$ (left) and $I=1$ (right).
    }\label{fig:gs_energy}
\end{figure*}

In Fig.\,\ref{fig:gs_energy}, we show the effective energies obtained from the correlators, $R(t)$ and $R^{\textrm{opt}}(t)$, together with the ground state energies obtained from the HAL QCD potential at the time slices $t/a=13,14,15$.
The effective energies from $R(t)$ and $R^{\textrm{opt}}(t)$ are almost identical, indicating that the R-correlator $R(t)$ dominantly overlaps with the ground state.
We also observe that the ground state energies from the LO potential coincide with the effective energies from the R-correlators within statistical fluctuations at large times $t/a = 13$-$16$ in both isospin channels.
This agreement indicates that the LO potential gives a reasonable description at low energies for the $\bar{D}N$ system. 

\bibliography{main}

\end{document}